# ION EXCHANGE IN SILICATE GLASS: MASS AVERAGE INTERDIFFUSION COEFFICIENT DETERMINATION.

**Guglielmo Macrelli (\*)**

(\*) Isoclima SpA – R&D Dept. Via A.Volta 14, 35042 Este (PD) Italy
guglielmomacrelli@hotmail.com

**Abstract**

The mass average interdiffusion coefficient $D_M$ is an approximated constant value of the interdiffusion coefficient which is relevant in the kinetics of ion exchange in silicate glasses. In this study it is presented a simple technique for its determination based on the weight change of a glass sample after ion exchange. The theoretical basis of the method is presented in detail together with the approximations assumed in considering the constancy of the $D_M$ . Experimental results are presented for soda-lime silicate glasses and it is demonstrated the correlation of $D_M$ with the ion exchange temperature following an Arrhenius type equation with a energy activation barrier and a pre-exponential factor. The determination of the mass average interdiffusion coefficient allows the estimation of the compression layer depth when a stress profile is build-up as a consequence of ion exchange. The estimated values of the compression layer depth have been compared with the ones measured by an optical technique based on differential surface refractometry (DSR). Values have been found quite compatible within the uncertainty limits indicated by the DSR method.

## I.  Introduction

Ion exchange in silicate glasses is an important process either scientifically and technologically. Apart from historical evidences of its applications[1], ion exchange has been systematically scientifically studied since the 1960's of last century[1,2]. Applications have been identified in glass strengthening[2,3,4] (chemical strengthening) and in optical waveguides[1,5]. Looking for a suitable definition we can set the following: Ion Exchange (IX) in silicate glasses is a non-equilibrium thermodynamic process between a glass and an ion source, it is driven by gradients of electrochemical potentials of relatively mobile network modifiers. This means that, to activate ion exchange, we need a glass matrix with relatively mobile ions (typically monovalent alkali) and an ion source with available mobile ions. The contact of the glass matrix with the ion source generates the ion exchange as a consequence of the gradients in the electrochemical potentials of the involved ionic species. During the ion exchange process the Si-O network of the glass may be assumed stable. Considering ion $A$ in the ion source (IS) and ion $B$ in the glass matrix (GM) the IX process may be schematically written as:

$$A_{IS} + B_{GM} \Leftrightarrow B_{IS} + A_{GM} \qquad (1)$$

and pictorially represented in Figure 1 where "$A$" is Potassium $K^+$ and "$B$" is Sodium $Na^+$.

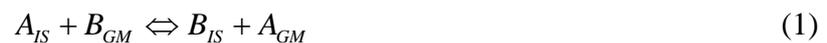

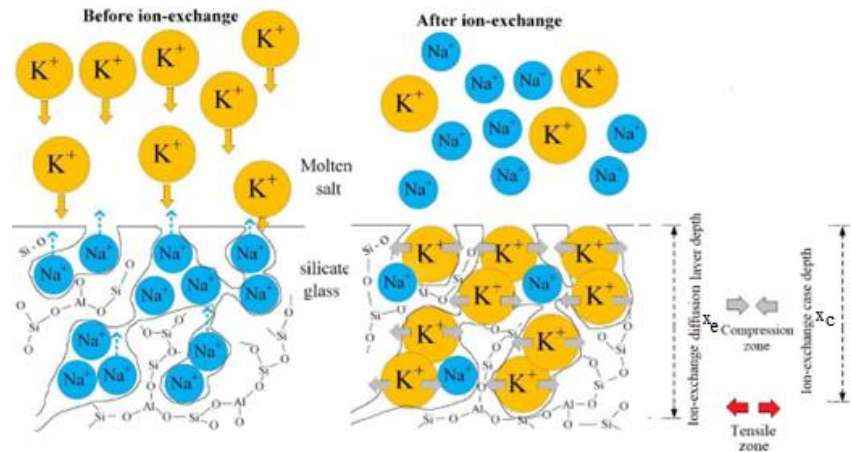

Figure 1 – Representation of IX process. On the right side they are indicated the layers of ions invasion and the concepts of diffusion layer depth $x_e$ and compression layer depth also named case depth $x_c$.

Typical assumptions for Ion Exchange in silicate glasses are:

(a) Mobilities of cations in the ion source are much higher than those in the glass matrix.
(b) Interface reactions between ion source and glass are so fast that the time to get an equilibrium condition at the interface is much lower than the overall contact time between ion source and glass matrix.

From the above assumptions it is clear that the process rate critical factor is the inter-diffusion of the involved ions in the glass matrix. It may happen that the two above conditions are not met, or not entirely met during real technological processes, this may occur in presence of ion source contaminations that reduce ion mobilities or induce interface blocking mechanism or when the IS/GM total ion exchange time is of the same order of the time needed to get a surface equilibrium condition. In Figure 2 the two processes are identified: an equilibrium process at the interface Source/Glass and an interdiffusion process in the glass body.

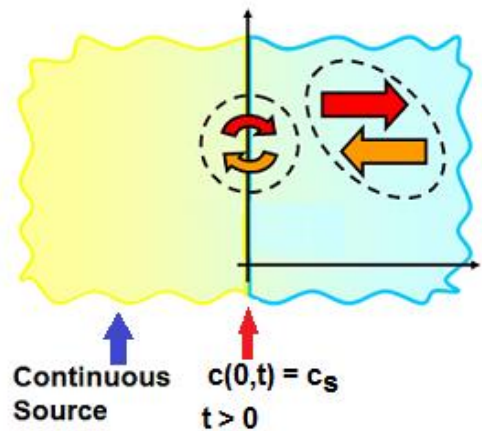

Figure 2 – Processes involved in ion exchange: surface equilibrium and interdiffusion in the bulk of material.

The main physical and chemical effects of Ion Exchange are, in a rational order:

(i) Near surface molar volume change when molar volume is different for the two exchanging ions
(ii) Near surface molar concentration change

(iii) Near surface refractive index change
(iv) Residual stress induction

Both refractive index profile and residual stress profile are a consequence of the concentration of the invading ions in the glass matrix. The refractive index profile is of relevance in optical applications (optical waveguides) and in residual stress measurements[4,6] (optical determination of residual stress by birefringence) while the stress profile is relevant in strengthening applications. In both cases a key role is played by the ion concentration of the invading ions as a consequence of the IX process. In summary to have ion exchange in silicate glass we need an alkali-containing silicate glass and the contact of the glass with a ion source of another alkali. These conditions activate the exchange of host ions for invading ions from the source (usually: molten salt bath, usually $KNO_3$). When ion exchange is applied for strengthening applications to generate a residual stress in the glass, temperature is managed to promote interdiffusion, and limit viscous relaxation. In this study we focus on the kinetics of the interdiffusion process which is characterized by the interdiffusion coefficient. The concept of mass average interdiffusion coefficient $D_M$ has been introduced by Varshneya and Milberg[7]. This parameter can be easily determined by measuring the change in weight of the glass article before and after ion exchange. The change in weight is due to the difference in molar weight between the incoming ion and the outgoing one. This method, known as weight gain or weight change[2,4], is just mentioned in an international standard[8] for glass chemical strengthening without entering in any detail about its implementation. The weight change method is reported by Bartolomew and Garfinkel[2] and by Gy[4] for the determination of the average interdiffusion coefficient and, although its evident simplicity, there aren't other systematic discussions and treatments. The theoretical basis of the weight gain method for the determination of the average interdiffusion coefficient are discussed in detail. It is also discussed a further development where, under some general assumptions, the depth of the compression layer can be estimated. An experimental example is presented where the weight gain method is used and the estimated compression layer depth is compared to the one measured by the differential surface refractometry (DSR) technique. Results are critically reviewed and discussed. In Table I the list of symbols, their definitions and units is reported.

## II. Theoretical part: Ion Flux Equations

The kinetics ion flux equations relevant to ion exchange are discussed by Macrelli, Mauro and Varshneya[9]. The relationship between ion fluxes $J_i$ and electrochemical potentials $\eta_i$ are established considering the components of the electrochemical potential, namely: the concentration related chemical potential $\mu_i$, the electrical potential (both external or internally generated because of different ion mobilities) $\varphi$, and the stress generated due to the different molar volumes of the involved ion species $\sigma_H$:

$$\eta_i = \mu_i + \bar{F}\varphi - V_i\sigma_H \quad , i=A,B, \tag{2a}$$

$$-J_i = C_i\beta_i \frac{\partial \eta_i}{\partial x} \quad . \tag{2b}$$

Applying electroneutrality and Gibbs-Duhem conditions, the individual flux equations for ion fluxes (2b) can be reduced[9] to a single Fick type equation:

$$-J_A = C_0 D_{AB} \left[ \left(1 + \frac{\partial \ln \gamma_A}{\partial \ln C_A}\right) + \frac{\alpha C_0}{RT} \chi_A (1-\chi_A) \Omega_{AB} \right] \frac{\partial \chi_A}{\partial x}. \tag{3}$$

In equation (3) concentration is expressed in terms of relative concentration:

$$C_A + C_B = C_0; \quad C_0 \chi_A + C_0 \chi_B = C_0; \quad \chi_i = \frac{C_i}{C_0}. \tag{4}$$

From (3) it appears evident the complexity of the interdiffusion coefficient, this is discussed in some details in the literature[9]. For the sake of this study and following a large part of literature[2,3,4,10] we take the approximated assumption of constant value for the interdiffusion coefficient $D_M$. After this assumption ion flux equation (3) is largely simplified in a more familiar first Fick equation between ion flux and concentration gradient:

$$-J_A(x,t) = D_M \frac{\partial C_A(x,t)}{\partial x}. \tag{5}$$

If the total contact time between the ion source and glass is $\tau$ and we indicate with $Q$ the total exchanged molar flux (mol/cm$^2$) we can derive some important relationships without making any additional assumption or approximation. To keep it more general, let's consider the first Fick equation (5) with a space and time variable diffusion coefficient and let's remove the subscripts $A$ and $M$. We make use of the Boltzmann transformation:

$$z = \frac{x}{\sqrt{D \cdot t}}, \tag{6}$$

In terms of the new variable $z$, the flux equation (5) results:

$$-J(z) = \sqrt{\frac{D}{t}} \frac{\partial C}{\partial z}. \tag{7}$$

From (7), the flux at the glass surface $x=0$ $(z=0)$ is:

$$-J_0(x=0,t) = D_0 \left[\frac{\partial C}{\partial x}\right]_{x=0} = \sqrt{\frac{D_0}{t}} \left[\frac{\partial C}{\partial z}\right]_{z=0}, \tag{8}$$

Where $D_0$ is the value of the diffusion coefficient at glass surface ($x=0$). The total exchanged molar flux $Q$ is just the integral over the total contact time (total time of ion exchange) $\tau$ of the ion flux at glass boundary:

$$Q = \int_0^\tau J_0(x,t)dt = -\sqrt{D_0}\left[\frac{\partial C}{\partial z}\right]_{z=0} \int_0^\tau \frac{1}{\sqrt{t}}dt = -2\sqrt{D_0 \cdot \tau}\left[\frac{\partial C}{\partial z}\right]_{z=0}. \tag{9}$$

Relationship (9) is very general and it is valid for any first Fick type equation even with variable diffusion coefficient. In our case we assume that the interdiffusion coefficient is constant and we can write the second Fick equation by setting the continuity condition:

$$\frac{\partial C_A}{\partial t} + \frac{\partial J_A}{\partial x} = 0, \qquad (10)$$

for (5) we obtain the diffusion equation (second Fick equation):

$$\frac{\partial C_A}{\partial t} = D_M \frac{\partial^2 C_A}{\partial x^2}. \qquad (11)$$

To solve this equation for the concentration of ion $A$ in the glass, we need an initial and a boundary condition. We can assume, without loosing any generality, that this equation is for the excess of ion $A$ in the glass, in this way the initial condition is $C_A(x,t)=0$. The boundary condition at glass/source interface can be set looking at Figure 2 and assuming an instantaneous equilibrium condition $C(x=0,t)=Cs$ for any $t \geq 0$. The conditions under which this boundary condition is reasonable have been discussed by Macrelli[11] and are substantially justified when the total ion exchange time is much larger than the time to attain kinetic equilibrium at the IS/GM interface. Under the initial and boundary conditions above indicated, the solution to the diffusion equation is[1,2,3,4,10]:

$$C_A(x,t) = C_s \, erfc\left(\frac{x}{2\sqrt{D_M t}}\right) = C_s \, erfc\left(\frac{z}{2}\right), \qquad (12a)$$

$$erfc(\zeta) = 1 - erf(\zeta) = 1 - \frac{2}{\pi}\int_0^\zeta \exp(-\beta^2)d\beta. \qquad (12b)$$

Taking this solution (12a), the derivative in (9) can be calculated:

$$\left[\frac{\partial C}{\partial z}\right]_{z=0} = \left[-\frac{C_s}{\sqrt{\pi}}\exp\left(-\frac{z^2}{2}\right)\right]_{z=0} = -\frac{C_s}{\sqrt{\pi}}. \qquad (13)$$

The total exchanged molar flux (9) results:

$$Q = 2C_s\sqrt{\frac{D_M \cdot \tau}{\pi}}. \qquad (14)$$

This result (14) for the total exchanged molar flux is coming from the kinetic flux equations under some approximated assumptions. On the other side the total exchanged molar flux is related to the weight change (weight gain) at final time $\tau$, $\Delta W(\tau)$ through a straightforward simple relationship:

$$Q = \frac{\Delta W(\tau)}{S_{ex}(M_A - M_B)}, \qquad (15)$$

where $S_{ex}$ is the exchange surface, and $M_A$ and $M_B$ are the molecular weights of the invading ions $A$ and of the outgoing ions $B$. Comparing equations (14) and (15) it finally results the expression for the average interdiffusion coefficient in term of the weight change:

$$D_M = \frac{\pi}{\tau}\left[\frac{\Delta W(\tau)}{2C_s \cdot S_{ex}(M_A - M_B)}\right]^2 \quad . \tag{16}$$

The determination of $D_M$ through the measurement of $\Delta W(\tau)$ allows the reconstruction of the concentration profile of the invading ions through equation (12a). Through the knowledge of the concentration profile other important information can be deducted through additional theoretical equations. In particular the residual stress (equi-biaxial stress) introduced in the glass is related to the residual concentration of the invading ions through equations that, in the simplest approximated case, do not take into account the stress relaxation of the glass matrix[1,2,3,10]:

$$\sigma(x,t) = -\frac{BE}{(1-\nu)}\left[C_A(x,t) - \overline{C_A(t)}\right] \tag{17}$$

Where, $E$ is the Young modulus of the glass, $\nu$ is the glass Poisson ratio and $B$ is the so called linear network dilatation coefficient (named also Cooper coefficient)[3,10] and $\overline{C_A(t)}$ is the average (over $x$) of the concentration:

$$\overline{C_A(x,t)} = \frac{1}{d}\int_0^d C_A(x,t)dx \quad , \tag{18}$$

and d is the glass thickness. From equation (17) we can argue that, as far as the invading ions concentration is higher than the average value, we have stress of compression (negative) while, if the concentration is lower than the average value, stress become positive indicating a tensile stress. The coordinate $x_c$ at which we have zero value of stress, corresponds to the so-called compression layer depth or case depth. The condition:

$$C_A(x_c,\tau) = \overline{C_A(\tau)} \quad , \tag{19}$$

set the zero value of residual stress and the solution of equation (19) for $x_c$ provides an estimate of the compression layer depth. According to the concentration profile (12a) the equation (19) is written:

$$erfc\left(\frac{x_c}{2\sqrt{D_M t}}\right) = \frac{\overline{C_A}}{C_s} \quad . \tag{20}$$

Average concentration of the invading ions can be evaluated by the change in weight:

$$\overline{C_A(\tau)} = \frac{\Delta W(\tau)}{V(M_A - M_B)}, \tag{21}$$

where $V$ is the glass sample volume. Another possible approach to evaluate average concentration is to calculate the average concentration directly from equations (19) and (12a):

$$\overline{C_A(\tau)} = \frac{C_s}{d}\int_0^d erfc(\frac{x}{\sqrt{D_M\tau}})dx = \frac{4C_s}{d}\sqrt{\frac{D_M\tau}{\pi}} \quad . \tag{22}$$

This last approach is acceptable when the glass thickness d is negligeable in comparison with the sample dimensions (ion exchange in glass edges is neglected in (22)). In general the approach of equation (21) is acceptable for any glass sample thickness.

Even though condition (19) for case depth has been derived for the stress equation (17) that do not consider relaxation effects, a similar argument can be used for the more general stress equation with relaxation effects[12,13]:

$$\sigma(x,t) = -\frac{BE}{(1-\nu)}\left[\left(\mathcal{V}(x,t)C_A(x,t) - \overline{\mathcal{V}C_A(t)}\right) - \int_0^t \frac{\partial R(t-t')}{\partial t'}\left[\mathcal{V}(x,t')C_A(x,t') - \overline{\mathcal{V}C_A(t')}\right]dt'\right], \quad (23)$$

where $\mathcal{V}(x,t)$ is the Varshneya relaxation function[12,13] and R(t) is the Maxwellian stretched relaxation function[3,4]. Condition (19) for equation (23) is

$$\mathcal{V}(x,\tau)C_A(x_c,\tau) = \overline{\mathcal{V}C_A(\tau)} \quad (24)$$

When the Varshneya function $\mathcal{V}(x,t)$ does not depend on the spatial coordinate than condition (24) relaxes to condition (19). For some specific glasses, namely a particular LithiumAluminoSilicate chemical composition[13], submitted to ion exchange, residual stress profile relaxes in a way that condition (24) is satisfied for at least two different points[13] on the *x* coordinate and, as the ion exchange process is prolonged, surface compression turn to tensile status.

In the following experimental part we use equation (16) for the determination of the mass average interdiffusion coefficient $D_M$ and condition (21) with equation (20) for the determination of the compression layer depth.

### III. Experimental part: Interdiffusion coefficient determination

In this study we refer to soda-lime silicate glass. The glass chemical composition has been determined by X-rays Fluorescence and it is reported in Table II. Samples with nominal dimensions 66mm x 66mm with a nominal thickness of 1.6mm have been prepared by cutting from a larger sheet. Particular care has been taken in the cutting process to avoid chipping at the edge corners. Glass samples have been used for three Ion Exchange experiments (indicated as Q1,Q2 and Q3) where two samples per experiment type have been exposed to a specific Ion exchange schedule (Temperature and Immersion time) according to Table II. The ion source is a bath of molten analytical grade Potassium Nitrate ($KNO_3$) manufactured by chemical synthesis. The experiments have been carried out in a range of temperatures within the melting point of the Potassium Nitrate (334°C) and the upper temperature at which the decomposition of the nitrate group $NO_3^-$ become significant (480°C). Samples have been carefully washed, cleaned and dried than weighed with a analytical calibrated balance with a sensitivity of 0.0001g (0.1 mg). They have been pre-heated for 15 minutes prior to immersion, at a temperature of 25°C lower than the salt bath temperature. At the end of the immersion cycle (in these experiments always 24 hours), samples have been removed from the salt bath allowed 5minutes dripping than cooled down to room temperature. After washing and cleaning and inspecting to detect any surface damage, they have been weighed again to determine weight change after ion exchange. Samples have been subsequently measured by differential surface refractometry[4,6,8] by using FSM 6000 instrument manufactured by Orihara-Japan Ltd[14]. The purpose of this last measurements is to determine compression layer depth $Cd$ (μm) and surface compression $S_C$ (MPa).

The FSM specification[14] indicates measurements uncertainties of +/-20MPa for surface compression and +/5μm for the compression layer depth.

Results of weight change test, namely the determination of the mass average interdiffusion coefficient according to equation (16) and the estimation of the compression layer depth according to equation (20), are reported in table IV. In equation (16) it is needed the value of surface concentration $C_s$ of Sodium ions and the exchange surface $S_{ex}$ of the sample exposed to ion exchange. The surface concentration of Sodium ions is determined by the value of the concentration fraction in weight of Sodium oxide $\chi_w(Na_2O)$, the density of the glass $\rho$ (2.484 g/cm$^3$) determined by the Archimede's method and the molecular weight of Sodium Oxide ($MW$ = 61.98 g/mol) according to:

$$C_S(Na) = \frac{2\rho\chi_w(Na_2O)}{MW(Na_2O)} r \quad . \tag{25}$$

The exchange surface is determined from the geometry of the sample (width, length and thickness) which is a prismatic body. The difference of the molecular weigh of the exchanging ions can be easily calculated from the individual values of the molecular weight of Potassium ($M_K$=39.0993 g/mol) and Sodium ($M_{Na}$=22.9898 g/mol).

In equation (25) the parameter $r$ represents the fraction of Sodium ions exchanged on the glass surface, if all Sodium is exchanged $r$ = 1 while if no Sodium is exchanged $r$ = 0. In all determinations of this study it has been assumed $r$ = 1.

## IV. Discussion.

From Table IV it is quite evident that the mass average interdiffusion coefficient increases with the ion exchange temperature. This means that it is reasonable to argue that the ion exchange kinetics rate is following similar mechanisms of temperature dependance of diffusion processes[3,10]. The temperature dependance of the mass average interdiffusion coefficient can be assumed to follow an Arrhenius type function with an activation energy (barrier) $\Delta H_d$ and a pre-exponential factor $D_0$ according to[3,10] the following equation:

$$D_M(T) = D_0 \exp\left[-\frac{H_d}{RT}\right]. \tag{26}$$

In equation (26) the activation barrier is expressed on a mole basis. The pre-exponential factor is interpreted as a parameter containing information about the entropy associated with the ion exchange interdiffusion reaction[10]. The experimental results of table IV are reported in Figure 3 on a logarithmic scale together with equation (26), best fitted with the least squares method to the experimental results. The values of activation energy barrier, $H_D$ and pre-exponential factor $D_0$ of the best fitting are:

$$H_d = 165.8 kJ/mol \quad ; \quad D_0 = 6.22 cm^2/s \tag{27}$$

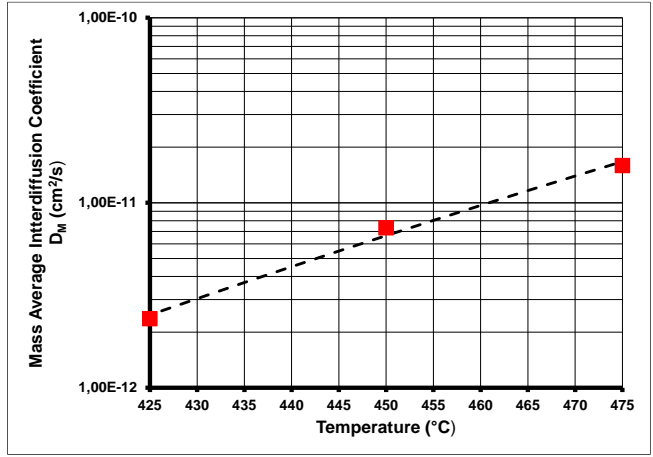

Figure 3 – Mass Average Interdiffusion coefficient $D_M$, red boxes are results of this study (Table IV), dashed curve is Arrhenius equation (26) with best fit parameters (27), correlation factor $R^2$=0.9924.

The activation barrier energy is compatible with the value of 150 kJ/mol reported by Gy[4] for soda-lime glass. The comparison of the compression layer depth estimated by the approach outlined in this study and formalized in equation (20), is compatible with the values measured by the DSR-FSM optical method. The difference ($x_c$-$Cd$) in the values of table IV is ranging from -2.4 to 2.1 µm that is 2 times lower than the declared uncertainty (+/- 5µm) of the FSM data sheet[14]. The weight change method proposed in this study provides an alternative and quite simple method to characterize ion exchanged silicate glasses. On the other side, because of the rough assumptions taken in this study, it is completely neglected the complexity of the interdiffusion coefficient in its dependence on concentration, stress and mutual interactions within the exchanging ions and with the silicate network[9]. It is important to emphasize the conceptual difference between the diffusion layer depth, indicated in figure 1 by the symbol $x_e$, and the compression layer depth indicated either as $x_c$ (Figure 1) and $c_d$. The first represents the characteristic depth of ions penetration in the glass matrix, it can be associated to the concept of diffusion length[15,16] or conventionally defined by a suitable convened low value of the concentration depth as proposed by Gy[4] ($c(x_e)$=0.005). The second represents the depth of the layer from the glass surface to the point where the compressive stress (see for example equation (17) is zero ($\sigma(x_c,t) = 0$).

## V. Conclusion

The weight change or weight gain measurement method outlined in this study is a quick and suitable approach to estimate the mass average interdiffusion coefficient $D_M$ and its dependency on ion exchange temperature. As a consequence of this determination the compression layer depth $x_c$ can be estimated as well. The simplicity of this method is based on rough approximations nevertheless it provides relevant information for industrial ion exchange process control and a guideline to more sophisticated approaches based on techniques for chemical depth profiling (secondary ion mass spectroscopy – SIMS, electron probe micro-analysis EMPA or x-ray photoelectron spectroscopy XPS) and the Boltzmann-Matano technique[16] suitably used when it is considered the concentration dependency of the interdiffusion coefficient.

# TABLES

## Table I – List of main symbols and units

| Symbol | Description | Units |
| --- | --- | --- |
| $M_A$ | Molecular weight of ion A | g/mol |
| $x,y,z$ | Spatial coordinates | m |
| $t$ | Time | s |
| $C_A(x,t)$ | Molar Concentration of ion A | mol/cm$^3$ |
| $J_A$ | Molar flux of ion A | mol/(cm$^2$s) |
| $Q_{AB}$ | Exchanged molar flux | mol/(cm$^2$) |
| $S_{ex}$ | Sample Surface through which we have ion exchange | cm$^2$ |
| $\Delta M_K$ | Molecular weight difference between ion species | g/mol |
| $D_M$ | Mass Average Inter Diffusion coefficient | cm$^2$/s |
| $T$ | Absolut Temperature | K |
| $R$ | Gas constant | J/(molK) |
| $\mathcal{F}$ | Faraday Constant | C/mol |
| $\eta$ | Electrochemical potential | J/mol |
| $\mu$ | Chemical potential | J/mol |
| $\beta_A$ | Mobility of Ion A | cm$^2$/J·s |
| $\gamma$ | Activity coefficient | Dimensionless |
| $x_c$, $Cd$ | Depth of compression layer, case depth | μm |
| $Sc$ | Surface Compression | MPa |
| $C_{rel}$ | Relative concentration | Dimensionless |
| $H_D$ | Activation Energy (Barrier) | J/mol |
| $D_0$ | Pre-exponential factor | cm$^2$/s |

**Table II – Chemical composition expressed in weight fraction $\chi_W$ and mole fraction $\chi_M$ (Determined by X-ray Fluorescence) of the silicate glass of this study (only main oxides are reported, minor oxides are not reported, their overall amount is less than 0.1 %).**

| Oxide | $\chi_W$ -Weight (%) | $\chi_M$ - Mol (%) |
|---|---|---|
| $SiO_2$ | 72.4 | 71.09 |
| $Al_2O_3$ | 0.55 | 0.32 |
| $Na_2O$ | 13.1 | 12.47 |
| $K_2O$ | 0.32 | 0.20 |
| $CaO$ | 9.02 | 9.49 |
| $MgO$ | 4.21 | 6.16 |

**Table III – Ion Exchange experiments of this study.**

| IX – Id. | Temperature (°C) | Immersion time (hours – h) |
|---|---|---|
| Q1 | 425 | 24 |
| Q2 | 450 | 24 |
| Q3 | 475 | 24 |

**Table IV – Results of Ion Exchange experiments of this study.**

| IX – Id. | Weight Change $\Delta W$ (g) | Mass Average Interdiffusion Coefficient from eq.(16) $D_M$ ($cm^2/s$) | Case depth from eq. (20) $x_c$ ($\mu m$) | Case depth measured by DSR – FSM $C_d$ ($\mu m$) | Surface Compression measured by DSR - FSM $S_c$ (MPa) |
|---|---|---|---|---|---|
| Q1 | 0.0079 | $2.365 \cdot 10^{-12}$ | 17.3 | 19.7 | 600 |
| Q2 | 0.0127 | $7.319 \cdot 10^{-12}$ | 28.5 | 28.7 | 477 |
| Q3 | 0.0206 | $1.589 \cdot 10^{-11}$ | 39.3 | 37.2 | 369 |